# Form of the exact partition function for the generalized Ising Model

T.R.S. Prasanna*

Department of Metallurgical Engineering and Materials Science

Indian Institute of Technology, Bombay

Mumbai 400076 India

The problem of N interacting spins on a lattice is equivalent to one of N clusters linked in a specific manner. The energy of any configuration of spins can be expressed in terms of the energy levels of this cluster. A new expression is obtained for the probability of occurrence of any configuration. A closed form expression is obtained for the partition function per spin in terms of the energy levels of this cluster with the degeneracies being a function of temperature. On physical grounds it is suggested to be the form of the exact partition function per spin. The partition functions of Ising-like systems all have a common form. For the 3D Ising model seven functions need to be determined to describe the partition function completely. The key to understanding second order phase transitions and critical phenomena lies in the temperature dependence of various degeneracies. It is necessary to develop new techniques to determine the partition function that account for this temperature dependence, as it would represent the underlying physics correctly.

The Ising Model can be considered to be the prototype for studies on second order phase transitions and critical phenomena. It has been the object of intense study since Onsager [1] developed the exact solution to the partition function of the two dimensional (2D) Ising model. The enormous interest is partly due to the simplicity in describing the model that captures many features of interest of second order phase transitions and critical phenomena. However, in spite of intense efforts over the last six decades, the exact solution to the partition function of the three dimensional Ising model or the generalized Ising Model – with interactions that are anisotropic, beyond the nearest neighbor or under applied field and/or with higher state spins– has remained elusive. Consequently, many approximations to the partition function have been developed. These approximations address the different interests of investigators, with some focused on critical phenomena and the determination of critical indices (series expansions [2], scaling [3-5], renormalization group theory [6,7] etc.) and others focused on developing partition functions valid over a larger temperature range (mean field theory, Bethe approximation [8], cluster variation methods [9], Monte Carlo simulations, series expansions etc.). These techniques are described in standard textbooks [10-14] and other specialized books [15-16]. While the series expansions give an infinite power series expression for the partition function, none of the approximations to date result in a closed form expression for the partition function valid over the entire temperature range of interest. A new approach to the Ising model is described in this paper. This method leads to a closed form expression to the partition function per spin and on physical grounds is suggested to be the form of the exact partition function for the generalized Ising Model.

The new approach differs fundamentally from previous approximations in that the starting point to determine the partition function is the energy of a single configuration and not the formal mathematical expression for the partition function, Q(T), given by

$$Q(T) = \sum_{s_1}\sum_{s_2}...\sum_{s_N} \exp((\sum_{<ij>} J_{ij} s_i s_j)/kT) \qquad (1)$$

Also, no attempt is made to determine the partition function as represented by eqn.1. For concreteness, the ideas presented here are developed with reference to the 2D Ising model under zero applied field condition without any loss of generality.

The energy of interaction between spins in the Ising model is given by $E_{ij} = -J_{ij} s_i s_j$. The energy of any configuration of N spins, $E_c$, is obtained by summing the energy of interaction, $E_{ij}$, over all pairs of spins while ensuring that double counting is avoided. However, it can also be determined in the following manner. The energy of interaction of each spin, $E_{cl}$, is defined by

$$E_{cl} = -\sum_{<ij>} J'_{ij} s_i s_j \qquad (2)$$

where $J'_{ij} = J_{ij}/2$ and the summation is over all spins interacting with it. The energy of any configuration of N spins is given by summing, over all spins, the energy of interaction of each spin

$$E_c = \sum_{n=1}^{N} E_{cl\,n} \qquad (3)$$

In this process, the interaction between two spins is double counted and results in the correct energy contribution of $-J_{ij} s_i s_j$. This process allows the energy of any configuration of N spins to be written as a sum of N terms, each of which is the energy ($E_{cl}$) of a cluster consisting of a particular spin and the spins with which it interacts with strength $J'_{ij}$. Because the various energy levels of the cluster are finite in number, this process shows that the energy of each of the N terms can only be one of a finite number of possible energy levels.

Fig. 1 illustrates the procedure for the 2D Ising model. The basic cluster consists of a central spin linked to four nearest neighbor spins it interacts with. As can be readily seen, the clusters are not independent and are linked in a specific manner. A cluster centered on spin (i′j′) links only to clusters centered on i′±1,j′, j′±1,i′ through a common "bond", on i′±2,j′, j′±2,i′ through a common vertex and on i′±1,j′±1 through a common "face". In this case, only five distinct energy levels result and the energy of any configuration of N spins can be written as a sum of N terms, each of which is one of the five distinct levels of the cluster.

A standard textbook [13] on Statistical Mechanics states in its introduction to Phase Transitions that "the characteristic feature of the interparticle interactions in these systems is that they *cannot* be 'removed' by means of a transformation of the coordinates of the problem; accordingly, the energy levels of the total system cannot, in any simple manner, be related to the energy levels of the individual constituents." The method described above shows that when systems that exhibit cooperative behavior are modeled as Ising systems, the difficulty described in the second part of the statement can be overcome. The energy of any configuration of N spins (the total system) can be represented in terms of the energy levels of a cluster. The problem of N interacting spins on a lattice can be transformed into a problem of N clusters of one particular type linked in a specific manner.

The relative weight of any configuration is given by the Boltzmann factor $e^{-E_c/kT}$ where $E_c$ is the energy of the configuration. Using eqn.3, this can be written as a product of N terms

$$e^{-E_c/kT} = \prod_{n=1}^{N} e^{-E_{cl\,n}/kT} \quad (4)$$

where $E_{cl\,n}$ is the energy of the cluster centered on spin n. Because of the extensive nature of the Helmholtz free energy, the partition function for a system of N spins can be written as

$$Q = q^N \quad (5)$$

which is a product of N terms, each of which is q – the partition function per interacting spin. Using eqns.4 and 5, the probability of occurrence of any configuration, $p_c$, can be written as a product of N terms

$$p_c = (e^{-E_c/kT})/Q = \prod_{n=1}^{N} (e^{-E_{cl\,n}/kT}/q) \quad (6)$$

This is a new expression and is strikingly similar to the probability of occurrence of any configuration of a system containing N independent identical objects (gas molecules, harmonic oscillators etc.)

The only energy levels that are present in the N exponential terms in eqn.6 are the energy levels $E_{cl}$, of the cluster. This is valid for any configuration. The structure of this expression shows that the only energy levels present in the partition function per spin (q) are those of the cluster. However, the degeneracies of the energy levels are unknown. Alternately, the problem can be viewed as one of linking N clusters of one particular type in a specific manner. Assuming the clusters to be independent leads to a ready determination of the partition function, $q_i$, of the independent cluster. Linking of clusters does not alter their energy levels but alters the degeneracies of various energy levels as some states become inaccessible. Both of the above arguments suggest that the partition function per interacting spin, q, can be written as

$$q = \sum_{j=1}^{M} b_j(T)(e^{-E_{cl_j}/kT}) \qquad (7)$$

where the summation is over all the energy levels, $E_{cl_j}$, of the cluster and $b_j(T)$ are their degeneracies. The partition function of a system of N interacting spins can be written as

$$Q = (\sum_{j=1}^{M} b_j(T)(e^{-E_{cl_j}/kT}))^N \qquad (8)$$

Eqn.7 shows that the partition function can be expressed in the form of a sum over energy levels, while hitherto only a sum over states representation of the partition function was available. This represents an alternate but equivalent framework to determine the partition function and shows that the partition functions of all Ising-like models have a common form. The entire effect of cooperative behavior is contained in the degeneracies. Eqn.7 reduces the problem of determining the partition function to that of determining the degeneracies, $b_js(T)$, of a finite number of energy levels.

The degeneracies in eqn.7 are shown to exhibit temperature dependence in the most general formulation. This is necessary on mathematical and physical grounds as discussed below. The partition function for the 2D Ising model was approximated by

eqn.7 with the assumption that the degeneracies, $b_j$s, were independent of temperature. The partition function per spin, q, was determined using both the high and low temperature series expansions. For the low temperature series expansion, the expression (eqn.15) for the reduced free energy per spin [17] in the thermodynamic limit (T = 700, 800, 900, 1000 and 1100 K) and for the high temperature series expansion the expression (eqn.6.7) given in Ref.12 was used (T = 3200, 3400, 3600, 3800 and 4000 K), with J/k = 700 K, to determine the partition function per spin, q. The values obtained were fit to the expression

$$q = (b_1\ e^{-2J/kT} + b_2\ e^{-J/kT} + b_3 + b_4\ e^{J/kT} + b_5\ e^{2J/kT}) \qquad (9)$$

with J = 2J′, obtained by expanding eqn.7 for this model under the assumption that the degeneracies are independent of temperature. This expression is simpler than the correct expansion of eqn.7 for the 2D Ising model given by

$$q = (b_1(T)\ e^{-2J/kT} + b_2(T)\ e^{-J/kT} + b_3(T) + b_4(T)\ e^{J/kT} + b_5(T)\ e^{2J/kT}) \qquad (10)$$

(Note that fitting to eqn.10 is much more difficult as the coefficients to be solved for are unknown functions). The resultant simultaneous equations obtained by fitting to eqn.9 were solved using Matlab 6.1. Table I shows that the values of the degeneracies obtained using high and low temperature series expansions are different. More importantly, negative values are obtained for the degeneracies of some energy levels, which is physically unacceptable. Moreover, assuming the degeneracies to be independent of temperature and of constant value leads only to a new independent cluster with different degeneracies and cannot represent an interacting spin system. The above observations suggest that the assumption of degeneracies having constant values leads to unacceptable consequences and hence the degeneracies are a function of temperature.

The second order phase transition results in singularities of various quantities at the critical point. This has been explained by the fact that an infinite sum of Boltzmann factors can result in a sum that is non-analytic [10,14]. However when expressed in closed form the expression for the partition function per spin (q) must contain the non-analyticity. The exact partition function per spin of the 2D Ising model exhibits this behavior. This suggests that in eqn.7, the various degeneracies, $b_j s(T)$, must contain the non-analyticity.

An important question that arises is whether the *exact* partition function per spin *can* be written in the form of eqn.7. Eqn.7 has been obtained using some of the fundamental ideas of Statistical Mechanics. The canonical ensemble partition function is the sum of Boltzmann factors over all configurations (states) and can be written equivalently in the form of a summation over energy levels as well. The first of the two arguments leading to eqn.7 is based on the fundamental idea that the only energy levels present in the partition function are those of the system (cluster). The second argument represents a physical fact that linking of clusters cannot alter its energy levels and can only alter its degeneracies. Based on these physical arguments, it is suggested that eqn.7 is the equivalent form of the exact partition function per spin. This form represents a rewriting of the partition function per spin as a sum over energy levels and can be applied to all lattice based models with discrete spin states including the generalized Ising model and the Potts model.

Eqn.7 sheds new light on the nature of the exact solutions to the 1D and 2D Ising models. The Hamiltonian for the Ising model is given in standard form[13] by

$$H = -\sum_{<ij>} J_{ij}\ s_i\ s_j \qquad (11)$$

where the summation is over all nearest neighbors. For the 1D Ising model, each spin has two nearest neighbors (resulting in clusters with energy levels $E_{cl}$ = -J, 0 and J) and eqn.7 for this model is given by

$$q = (b_1(T)\ e^{-J/kT} + b_2(T) + b_3(T)\ e^{J/kT}) \qquad (12)$$

The exact result for the 1D Ising model[10-13] is

$$q = 2\cosh J/kT = e^{-J/kT} + e^{J/kT} \qquad (13)$$

A comparison with eqn.12 shows that the exact solution for the 1D Ising model is in the *form* of eqn.7. Among the possible sets of values of the various degeneracies are A) $b_1(T) = b_3(T) = 1$, $b_2(T) = 0$ and B) $b_1(T) = (\frac{1}{2}\ e^{2J/kT})$, $b_2(T) = (\frac{1}{2}\ e^{J/kT} + \frac{1}{2}\ e^{-J/kT})$ and $b_3(T) = (\frac{1}{2}\ e^{-2J/kT})$. Possibility A can be ruled out on physical grounds because all the degeneracies have constant values which implies that the clusters are independent. Possibility B is a viable solution because the degeneracies are functions of temperature as is suggested to be for a system of linked clusters. However, the degeneracies are analytic and hence the 1D Ising model does not result in a phase transition. There could be many other solutions and the exact functional form of the degeneracies must be considered to be unknown at this point.

Eqn.10 represents the expansion of eqn.7 for the 2D Ising model. The exact partition function for the 2D Ising model can be written as[10,11]

$$q = 2\cosh(2J/kT)e^I \qquad (14)$$

where I is given by

$$I = \tfrac{1}{2\pi}\int_0^\pi d\phi \ln\{\tfrac{1}{2}[1+(1-\kappa^2\sin^2\phi)^{1/2}]\} \tag{15}$$

with

$$\kappa = 2\sinh(2J/kT)/\cosh^2(2J/kT) \tag{16}$$

Expanding cosh(2J/kT) in eqn.14 and comparing with eqn.10 suggests the simplest assignment for the various degeneracies are F) $b_1(T) = b_5(T) = e^I$ and $b_2(T) = b_3(T) = b_4(T) = 0$. Another possible assignment is G) $b_1(T) = \tfrac{1}{2} e^I$, $b_2(T) = \tfrac{1}{4} e^{-J/kT} e^I$, $b_3(T) = \tfrac{1}{4} (e^{-2J/kT} + e^{2J/kT}) e^I$, $b_4(T) = \tfrac{1}{4} e^{J/kT} e^I$ and $b_5(T) = \tfrac{1}{2} e^I$.

The degeneracies $b_1(T) = b_5(T) = e^I$ are temperature dependent and non-analytic in assignment F and this confirms the suggestion that the degeneracies must contain the non-analyticity for systems to exhibit critical behaviour. However, the degeneracies $b_2(T)$, $b_3(T)$ and $b_4(T)$ (corresponding to energies $E_{cl}$ = -J, 0 and J) are identically zero. This is problematic because the probability of occurrence (eqn.6) of the vast majority of configurations at finite temperatures will contain clusters with energies of –J, 0 and J but the denominator, which is the partition function (eqn.10), under assignment F does not contain terms corresponding to these energies. Hence, assignment F is unacceptable.

Even though the exact partition function (eqn.14) for the 2D Ising model has been known for sixty years, eqn.6, a new expression, shows that the understanding is incomplete. The probability of occurrence of any configuration can *always* be written in the form of eqn.6 at all temperatures. The numerator contains energy levels $E_{cl}$ = -2J, -J, 0, J and 2J only and even using the exact partition function (eqn.14) for the denominator does not help in identifying the contribution of the energy levels $E_{cl}$ = -J, 0, J to the partition function. One explanation is that the exact partition function is the resultant sum of various terms that contain the contribution of individual energy levels. However, by summing the contributions of individual energy levels the underlying physics is obscured.

While speculative, assignment G for the various degeneracies when substituted in eqn.10 sums up to the exact result and also shows the contribution of each energy level to the partition function explicitly. This would allow the expression for the probability of occurrence (eqn.6) to be understood. However, the exact functional form of the degeneracies for the 2D Ising model (as for the 1D Ising model) must be considered to be unknown currently.

From the above analysis, it is clear that the understanding of even a well-studied system such as the 2D Ising model is incomplete. Developing new solution methods that determine the partition function *and* provide additional information about the functional form of the degeneracies of various energy levels are necessary. This information is crucial as the temperature dependence of the degeneracies contain the essential physics underlying phase transitions and critical phenomena. It would also provide a transparent understanding of the expression for the probability of occurrence of any configuration (eqn.6). The additional information would shed light on whether *all* or only *some* of the degeneracies are non-analytic. Also, if the various degeneracies are given by *different* non-analytic functions, there must exist some relationship amongst them to ensure that the singularities of various thermodynamic quantities occur at only one temperature, the critical temperature. These new solution methods would include approximate methods where the approximation would be for the functional form of the degeneracies.

Despite intense efforts over the past sixty years, the exact partition function (as a single term expression) for the 3D Ising model (6 nn) has remained elusive. Eqn.7 shows that only seven functions, $b_1(T) \ldots b_7(T)$, need to be determined because it gives the partition function per spin to be

$$q = (b_1(T)\, e^{-3J/kT} + b_2(T)\, e^{-2J/kT} + b_3(T)\, e^{-J/kT} + b_4(T) + \\ b_5(T)\, e^{J/kT} + b_6(T)\, e^{2J/kT} + b_7(T)\, e^{3J/kT}) \tag{17}$$

This raises a new possibility that it may be possible to determine the exact partition function as a sum of seven terms but that it may not be possible to sum these terms to obtain a single term expression.

Finally, determining the partition function in the form of eqn.7 appears disadvantageous, as it requires determining a finite number of functions (degeneracies) as opposed to one function of temperature (partition function). However, it is a new, alternate and equivalent representation of the partition function and some of the reasons to fully explore this approach are a) It is necessary to do so for a complete understanding of the system. If the partition function were to be determined as a single term, a simple question can be posed for which no answer is available, e.g. contribution of the various energy levels of the cluster to the partition function b) It reflects the underlying commonality in the structure of partition functions for Ising-like models while a single term expression does not c) Eqn.7 raises a new possibility that it may be possible to determine the exact partition function as a sum of finite terms, but it may not be possible to sum these terms to obtain a single term expression. d) Because for the 3D and more complicated Ising-like models approximations are the only recourse left and eqn.7 gives the form of the exact partition function, it follows naturally that new approximations should be developed that at least retain this form. e) Unless it can be shown that the ideas and results obtained from this (sum over energy levels) framework are in direct correspondence with those obtained from the previous (sum over states) framework, it will yield new insights into the nature of phase transitions and critical phenomena.

In conclusion, this paper describes a simple method to obtain the form of the exact partition function for Ising-like models. This represents an alternate and equivalent framework to determine the partition function. It is necessary to investigate further to fully estimate the possibilities of this approach towards understanding phase transitions and critical phenomena.

email: prasanna@met.iitb.ac.in

Table I  Degeneracies ($b_j$) of the various energy levels of the cluster determined by fitting the partition function (9) to that obtained from high and low temperature series expansions

|        | $b_1$ | $b_2$   | $b_3$  | $b_4$  | $b_5$ |
|--------|-------|---------|--------|--------|-------|
| Low T  | 20.13 | -32.53  | 19.99  | -5.50  | 2.57  |
| High T | 3076  | -10120  | 12487  | -6487  | 1408  |

# Figures

1. Any configuration of N spins (for the 2D Ising Model) is equivalent to N linked clusters. The basic cluster consists of a central spin and the neighboring spins interacting with strength $J' = J/2$. Double lines indicate that only one line is to be considered when calculating the energy of cluster centered on one spin and the other line is considered when calculating the energy of the cluster centered on the neighboring spin. The possible energy levels of this cluster are $-4J'$ $(-2J)$, $-2J'$ $(-J)$, $0$, $2J'(J)$ and $4J'$ $(2J)$. Each cluster links to other clusters through common "faces"(dash dot lines), "bonds"(dashed lines) and vertices (dotted lines). Diagonal lines mark the boundaries of the clusters and are drawn for illustrative purposes.

Figure1. T.R.S. Prasanna

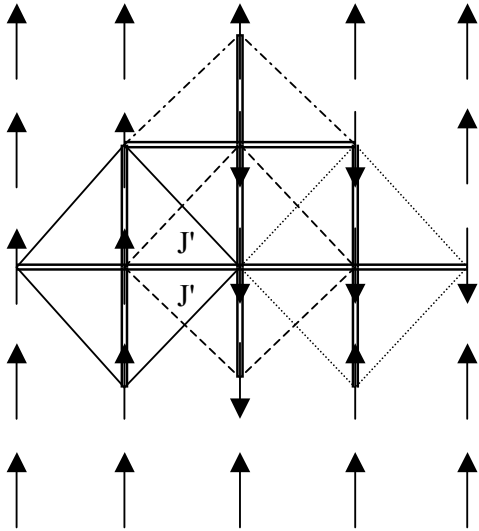